\documentstyle[prc,aps,manuscript,epsfig]{revtex}

\begin{document}
 \newcommand{\df}{\, d}
\title{Semimicroscopical description of the simplest photonuclear reactions
accompanied by excitation of the giant dipole resonance in medium-heavy mass nuclei}
\author{V.A. Rodin$^{1,2}$ and M.H. Urin$^{2}$}
\address{1. Institut f\"ur Theoretische Physik der Universit\"at T\"ubingen,\\
   Auf der Morgenstelle 14, D-72076 T\"ubingen, Germany\\
2. Department of Theoretical Nuclear Physics,\\
Moscow Engineering and Physics Institute (State University),\\ 115409 Moscow, Russia}
 \maketitle
 \begin{abstract}
A semimicroscopical approach is applied to describe photoabsorption and partial
photonucleon reactions accompanied by the excitation of the giant dipole resonance
(GDR). The approach is based on the continuum-RPA (CRPA) with a
phenomenological description for the spreading effect. The phenomenological
isoscalar part of the nuclear mean field, momentum-independent Landau-Migdal
particle-hole interaction, and separable momentum-dependent forces are
used as input quantities for the CRPA calculations. The experimental photoabsorption
and partial $(n,\gamma)$-reaction cross sections in the vicinity of the GDR are
satisfactorily described for $^{89}$Y, $^{140}$Ce and $^{208}$Pb target nuclei.
The total direct-neutron-decay branching ratio for the GDR in $^{48}$Ca and
$^{208}$Pb is also evaluated.

\end{abstract}

\section{Introduction}

A detailed description of the familiar (isovector, $T_3=0$) giant dipole resonance in
medium-heavy mass nuclei is a long-standing theoretical problem. A number of
macroscopical, semiclassical and microscopical approaches have been used to describe
the photoabsorption cross section, the latter being proportional to the energy-weighted
isovector dipole strength function \cite{dan95}. CRPA-based microscopical approaches have been developed
during the last two decades to describe the E1 strength function \cite{auer83}--\cite{mur94}.
However, partial photonucleon reactions accompanied by the GDR excitation have not been
studied in these approaches. Experimental data on these reactions (mainly on the
neutron radiative capture) are usually described within the DSD-model (the model of
direct+semidirect capture). The model was originally proposed by Brown~\cite{brown} and
substantially extended by the Lublyana group (see e.g. Ref.~\cite{lik98} and references
therein). Within the model a number of phenomenological quantities are used to
parameterize the energy-averaged reaction-amplitude and,
in particular, a rather large imaginary part of the GDR form factor
is used without a clear physical understanding of the origin of this part.

Partial photonuclear reactions accompanied by
GDR excitation are closely related to GDR direct decay into the respective
nucleon channels and carry information on the GDR structure.
Therefore, the CRPA-based microscopical approaches, which are
able to describe direct nucleon decays of the GDR (see e.g. Ref.~\cite{mur94}), can be extended
to describe the partial photonucleon reactions.

An attempt to use the CRPA-based
semimicroscopical approach to describe partial photonucleon reactions accompanied
by the GDR excitation was undertaken in Ref.~\cite{Chek95}. Later the approach was
extended and applied by Urin and coworkers for describing
the direct-nucleon-decay properties of a number of charge-exchange (isovector) and
isoscalar giant resonances (see Refs.~\cite{Moukh99} and \cite{Gor00}, respectively).
The ingredients of the semimicroscopical approach are the following:
\begin{itemize}
\item[(1)] the phenomenological isoscalar part of the nuclear
mean field (including the spin-orbit term) and the momentum-independent Landau-Migdal
particle-hole interaction, together with some partial self-consistency conditions;
\item[(2)] a phenomenological account for the spreading effect (that is due to coupling of
particle-hole-type doorway states to many-quasiparticle configurations) in terms of
an energy-dependent smearing parameter.
\end{itemize}

However, it was found in Ref.~\cite{Chek95} that the GDR energy and also the photoabsorption
cross section integrated over the GDR region, $\sigma^{int}_{GDR}$, are noticeably
underestimated in the calculations as compared with the respective experimental values~\cite{Ber75}.
The experimental value of $\sigma^{int}_{GDR}$, which exceeds the TRK sum rule,
$\sigma^{int}_{TRK}=60\,\displaystyle\frac{NZ}{A}$  MeV$\cdot$mb,
can be reproduced in selfconsistent calculations provided momentum-dependent forces are taken into
account. In Ref.~\cite{Rod00} we provided a satisfactory description of the
 experimental GDR energy in $^{208}$Pb and some partial
$^{208}$Pb$(n,\gamma)$-reaction cross sections in the GDR region by incorporating separable
isovector momentum-dependent forces with the intensity normalized to describe the
experimental $\sigma^{int}_{GDR}$ value and without the use
of additional adjustable parameters.

In the present work we extend the approach of Ref.~\cite{Rod00} as follows.
First, we use a simplier phenomenological realization for the spreading effect by
introducing an energy-dependent smearing parameter directly into
the CRPA equations.
Secondly, we take into consideration also the isoscalar part of
momentum-dependent forces in terms of an effective nucleon mass.
Thirdly, the photonucleon-reaction cross sections are calculated for a number of nuclei
and the results are compared with available experimental data. We make also
predictions for several partial $(\gamma,n)$-reaction
cross sections in the vicinity of the GDR in $^{208}$Pb and,
consequently, calculate partial direct-neutron decay branching ratios for this GR.
The total branching ratio is also evaluated for the GDR in $^{48}{\rm Ca}$
and compared with recent $(e,e'n)$-reaction data \cite{Str00}.

\section{Basic equations}

The CRPA equations are given below in the form adopted from the Migdal's finite-
Fermi-system theory~\cite{mig}. Similar to
Refs.~\cite{Chek95},\cite{Rod00}, the isobaric structure
of the GDR can be simplified by using the
quantity $D_z=\sum_a d_z(a)$, with $d_z=-\frac12\tau^{(3)}z$
as the z-projection of the dipole operator in the limit $(N-Z)/A\ll 1$.
The photoabsorption cross section, $\sigma_a(\omega)$
($\omega$ is the gamma-quantum energy), is then
proportional to the dipole strength function $S(\omega)$:
\begin{equation}
\sigma_a(\omega)=B \omega S(\omega)
\label{sigm_a}
\end{equation}
with $B=4\pi^2\frac{e^2}{\hbar c}$~.
Within the CRPA the strength function is determined
by the effective dipole operator $\tilde d_z(\omega)$, which differs from $d_z$ due
to polarization effects caused by the particle-hole interaction
$\hat F=\frac1{2} \displaystyle\sum\limits_{1\neq2} F(1,2)$.
Together with the momentum-independent isovector part of the Landau-Migdal
interaction, $F_{L-M}(1,2)=F'\cdot (\vec \tau_1 \vec \tau_2) \delta (\vec r_1 - \vec r_2)$,
we also use translationally invariant separable momentum-dependent forces:
\begin{equation}
\hat F_{m-d}=-\frac{1}{4mA} \sum_{1,2}(\kappa_0+\kappa'
(\vec \tau_1 \vec \tau_2 ))(\vec p_1-\vec p_2)^2\simeq -\frac{\kappa_0}{2m}
\sum_{1} \vec p_1^2 +\frac{\kappa'}{2mA}\sum_{1,2}(\vec \tau_1 \vec \tau_2 )
(\vec p_1 \vec p_2)\,,
\label{F12}
\end{equation}
where, $m$ is the nucleon mass, $A$ is the number of nucleons, and $\kappa_0$ and $\kappa'$
are the
isoscalar and isovector strengths of the momentum-dependent forces, respectively.
The approximate equality in Eq.~(\ref{F12}) is valid provided that the isovector $1^-$
excitations are analyzed within the limit $(N-Z)/A\ll 1$.

The use of momentum-dependent forces approximated by
Eq.~(\ref{F12}) allows one to:
\begin{itemize}
\item[(1)]
obtain simple expressions for the effective mass,
$m^*=m/(1-\kappa_0)$, and also for the
sum rule $\sigma^{int}=\int \sigma_a(\omega) d\omega=(1-\kappa_0)(1+\kappa')\sigma^{int}_{TRK}$;
\item[(2)] use the following expression for $\tilde d_z(\omega)$:
\begin{equation}
\tilde d_z(\omega)=-\frac12\tau^{(3)}\left(V(r;\omega) \frac zr+
\frac{i}{\hbar}\Delta(\omega)p_z\right).
\end{equation}
\end{itemize}
Having separated isobaric and spin-angular variables in the CRPA equations, the
following expressions for $S(\omega)$ and the
components of the effective dipole operator within the above-mentioned approximation
can be obtained:
\begin{equation}
S(\omega)=-\frac{2}{3} Im \int
r \left[A(r,r';\omega)+A_\kappa(r,r';\omega)\right]
V(r';\omega)\df r\df r'\ ,
\label{strfun}
\end{equation}
\begin{equation}
A_\kappa(r,r';\omega)=-\frac{\kappa \int A(r,r';\omega)\hat L(r')\df r'
\int \hat L(r)A(r,r';\omega)\df r}
{1+\kappa\int \hat L(r) A(r,r';\omega) \hat L(r')\df r\df r'} \ ,
\end{equation}
\begin{equation}
V(r;\omega)=r+
\frac{2F'}{r^2}\int \left[A(r,r';\omega)+A_\kappa(r,r';\omega)\right]
V(r';\omega)\df r'\ ,
\end{equation}
\begin{equation}
\Delta(\omega)=-\frac{\kappa \int \hat L(r)A(r,r';\omega)V(r';\omega)\df r\df r'}
{1+\kappa\int \hat L(r) A(r,r';\omega) \hat L(r')\df r\df r'}\ .
\end{equation}
where, $\kappa=\displaystyle\frac{8\pi\hbar^2\kappa'}{3mA}$, and
the operator ${\hat L}(r)$ is defined in terms of spin-angular matrix elements
as follows:
$\langle(\lambda)\Vert{\hat L}(r)\Vert
(\lambda')\rangle=\displaystyle{\partial
\over{\partial r}}+\displaystyle{B_{(\lambda)(\lambda')}\over{r}}$
with $(\lambda)=\{j_\lambda, l_\lambda\}$, $B_{(\lambda')(\lambda)}=
(l_\lambda(l_\lambda+1)-l_{\lambda '}(l_{\lambda '}+1))/2$.
Furthermore, $A(r,r';\omega)=\frac12(A^n+A^p)$, where
$(rr')^{-2}A^\alpha(r,r';\omega)$ is the radial part
of the free particle-hole propagator carrying the GDR quantum numbers
($\alpha=n,p$). The propagators $A^\alpha$ can be presented in the
following form~:
\begin{eqnarray}
&A^\alpha(r,r';\omega)=\sum_{\mu,(\lambda)} (t^{(1)}_{(\lambda)(\mu)})^2
n^\alpha_\mu [\chi^\alpha_\mu(r)
g^\alpha_{(\lambda)}(r,r';\varepsilon_\mu+\omega)\chi^\alpha_\mu(r')
+
\label{propag}\\
&\chi^\alpha_\mu(r')
g^\alpha_{(\lambda)}(r',r;\varepsilon_\mu-\omega)\chi^\alpha_\mu(r)],
\nonumber
\end{eqnarray}
where $\mu=\{\varepsilon_\mu,(\mu)\}$
is the set of quantum numbers for single-particle
bound states;
$t^{(L)}_{(\lambda)(\mu)}=
\langle (\lambda) \Vert Y_L \Vert (\mu) \rangle \big/\sqrt{2L+1}$
is the kinematic factor (the definition
of the reduced matrix elements $\langle (\lambda)
\Vert Y_L \Vert (\mu) \rangle$ is taken in accordance with Ref.~\cite{W_E}),
$n_\mu=N_\mu/(2j_\mu+1)$ is the occupation factor ($N_\mu$
is the number of nucleons occupying level $\mu$);
$r^{-1}\chi^\alpha_\mu(r)$ is the bound-state radial wave function;
$(rr')^{-1}g^\alpha_{(\lambda)}(r,r',\omega)$ is the Green's function of
the single-particle radial Schr\"odinger equation.

A realization of the optical theorem follows from Eqs.~(\ref{sigm_a})-(\ref{propag}):
\begin{equation}
\sigma_a(\omega)=\sum_{\mu(\lambda)\alpha}
\sigma_{\mu(\lambda)\alpha}(\omega)\; ;\ \ \
\sigma_{\mu(\lambda)\alpha}(\omega)=
\left|M_{\mu(\lambda)\alpha}(\omega)\right|^2\; ;
\label{optth}
\end{equation}
\begin{equation}
\hskip-5mm M_{\mu(\lambda)\alpha}(\omega)=
\left(\frac{\pi}{3} B\omega n^\alpha_\mu\right)^{1/2}
t^{(1)}_{(\lambda)(\mu)}\int\chi^{(+)\alpha}_{\varepsilon(\lambda)}(r)
(V(r,\omega)+\Delta(\omega)\hat L(r)) \chi^\alpha_\mu(r)\df r~.
\label{ampl}
\end{equation}
In Eq.~(\ref{optth}), $\sigma_c$ is the double partial $(\gamma ,N)$-reaction cross section,
and $c=\{\mu(\lambda)\alpha\}$ is a set of channel quantum numbers.
Therefore, $\sigma_{\mu\alpha}(\omega)=\sum_{(\lambda)}
\sigma_{\mu(\lambda)\alpha}(\omega)$ is the partial cross section
corresponding to the population of the one-hole state $\mu$ in the
$\alpha$-subsystem of the product nucleus (only nuclei without
nucleon pairing
are considered).
In Eq.~(\ref{ampl}), $r^{-1}\chi^{(+)\alpha}_{\varepsilon(\lambda)}$ is the radial
scattering wave function (normalized to the $\delta$-function of energy) with
the escape-nucleon energy given by $\varepsilon=\varepsilon_\mu^\alpha+\omega$.

The formula for the cross section of the
inverse reaction (nucleon radiative capture by closed-shell
nuclei) is derived using the detailed balance
principle:
\begin{equation}
\sigma_c^{inv}(\omega)=(\omega^2/2mc^2\varepsilon)\sigma_{c}(\omega)
\;  ;\ \ \
\sigma^{inv}_{\mu\alpha}=
\sum_{(\lambda)}\sigma_{\mu(\lambda)\alpha}^{inv},
\label{inv}
\end{equation}
where $\varepsilon$ is the kinetic energy of the captured nucleon.
In deriving these formulae (in which both $\sigma_c$'s correspond to
the same compound nucleus) we neglect the difference between the effective
dipole operators calculated for the $A$ and $A+1$ nuclei: this
assumption is expected to be valid with an accuracy of
$A^{-2/3}$. Also, we note that the quantities
$\sigma_{\mu(\lambda)\alpha}$ should be
calculated by setting $n_\mu=1$ in Eq.~(\ref{ampl}), as follows
from the consideration of kinematics of the $(n,\gamma)$
reaction on closed-shell target nuclei.

The amplitudes $ M_{\mu(\lambda)\alpha}$ in Eq.~(\ref{ampl}) also
determine the anisotropy parameters, $a_{\mu\alpha}$,
in the gamma-quantum angular distribution:
$4\pi\displaystyle\frac{d\sigma^{inv}_{\mu\alpha}}
{d\Omega}(\varepsilon,\theta)=
\sigma^{inv}_{\mu\alpha}(\varepsilon)(1+a_{\mu\alpha}P_2(\theta))$,
where $P_2$ is the second degree Legendre polynomial. The expression
for $a_{\mu\alpha}$ can be presented in the form:
\begin{equation}
a_{\mu\alpha}=-\sqrt{30\pi}\sum_{(\lambda)(\lambda')}
i^{l_\lambda-l_{\lambda'}}
W(2j_\lambda'1j_\mu;j_\lambda 1)t^{(2)}_{(\lambda)(\lambda')}
(M_{c'})^* M_{c}/\sum_{(\lambda)}\vert  M_{c} \vert^2 \ ,
\label{a2}
\end{equation}
where $W(abcd,ef)$ is a Racah coefficient and $c'=\alpha,\mu,(\lambda')$.

We emphasize that the above-mentioned expressions for the reaction amplitudes and cross
sections are obtained within the CRPA. To calculate the energy-averaged amplitudes
accounting for the spreading effect, we solve
Eqs.~(\ref{strfun})-(\ref{propag}) and (\ref{ampl})
with the replacement of $\omega$ by $\omega+\frac i2 I(\omega)$. The
form of the smearing parameter $I(\omega)$
(the mean doorway-state spreading width) is taken to be
similar to that obtained
for the imaginary part of the nucleon-nucleus potential in some versions of the optical
model (see e.g. Ref.~\cite{mah}), namely
 \begin{equation}
I(\omega)=\alpha (\omega - \Delta)^2/[1+(\omega - \Delta)^2/B^2]\ ,
 \label{2}
 \end{equation}
where $\alpha, \Delta, B$ are adjustable parameters. A reasonable description of the
GR total width was obtained in Refs.~\cite{Moukh99,Gor00},
and also in Ref.~\cite{Rod00} by
using the parameterization given by Eq.~(\ref{2}). The approach
described above allows one
to calculate the energy-averaged photoabsorption cross section
$\bar \sigma_a(\omega)$ and reaction amplitudes $\bar M_c$. The partial
photonucleon-reaction cross sections $\bar \sigma_{\mu \alpha}$,
$\bar \sigma_{\mu \alpha}^{inv}$ and anysotropy parameters $\bar a_{\mu \alpha}$
are determined by the averaged amplitudes in the same way, as given by Eqs.
(\ref{ampl}),(\ref{inv}),(\ref{a2}), provided that the fluctuational part of the cross sections is neglected.
Cross sections $\bar \sigma_{\mu \alpha}(\omega)$ determine the partial direct-nucleon-decay
branching ratios, $b_{\mu \alpha}$, for the GDR according to the relation:
\begin{equation}
b_{\mu\alpha}=\int \bar \sigma_{\mu\alpha}(\omega)d\omega \left/
\int \bar \sigma_a(\omega)d\omega\ , \right.
\label{eq1}
\end{equation}
where integration is performed over the GDR region.

\section{Calculational ingredients and results}

The experimental data on partial photonucleon reactions accompanied by
the GDR excitation in medium-heavy mass nuclei
are very scarce. Two sets of the experimental data on the neutron radiative capture with the
GDR excitation in $^{208}$Pb~\cite{Berg72}
and in $^{89}$Y, $^{140}$Ce~\cite{Berg78} are only available.
Before going to the results of the
semimicroscopical description of these and some other data, we would like to comment
on the ingredients of the approach and the smearing procedure.

As in Refs.~\cite{Moukh99,Gor00}, the nuclear mean field is taken as the sum of the isoscalar
(including the spin-orbit term), isovector and Coulomb terms:
\begin{equation}
U(x)=U_0(r)+U_{so}(r){\vec \sigma} {\vec l}+
\frac 1{2} \tau^{(3)}v(r)+\frac 1{2} (1-\tau^{(3)})U_c(r)\,,
\label{pot}
\end{equation}
with $U_0(r)=-U_0 f_{WS}(r,R,a)$ and
$U_{so}(r)=-U_{so} df_{WS}(r)/rdr $, where
$f_{WS}$ is the Woods-Saxon function with $R=r_0 A^{1/3}$, $r_0=1.24$ fm,
$a=0.63$ fm, and $U_{so}=13.9(1+2(N-Z)/A)$ MeV fm$^2$.
The symmetry potential $v(r)$ in Eq.~(\ref{pot}) is calculated in a self-consistent way:
$v(r)=2F'\rho^{(-)}$, where $\rho^{(-)}=\rho^n-\rho^p$ is the neutron excess density,
$F'=300 f'$ MeV fm$^3$ with the Landau-Migdal parameter $f'=1.0$.
The Coulomb part in Eq.~(\ref{pot}) is calculated in the Hartree approximation
via the proton density $\rho^p$. The isoscalar mean-field depth $U_0$ is chosen to describe
experimental nucleon separation energies in the nuclei under consideration and
depends on the choice of the effective mass [or parameter $\kappa_0$ in Eq. (2)].
In our calculations we used $m^*=m$ and also the "realistic" value
$m^*=0.9\,m$. The extracted values of $U_0$
are listed in Table~1. The proton pairing in $^{89}$Y and $^{140}$Ce is
approximately taken into account within the BCS model to calculate
only the proton separation energy. In these calculations the pairing
gap is determined from the experimental pairing energies.

To apply the smearing procedure, we calculate the $\omega$-dependent
single-particle quantities in Eqs.~(\ref{propag}) and (\ref{ampl})
[Green's functions and continuum-state wave functions] using the single-particle potential
$U(x)\mp \frac i2 I(\omega)f_{WS}(r,R^*,a)$. The cut-off radius is chosen as
$R^*=2R$ in calculations of the Green's functions (i.e., in calculations of the effective dipole operator
and, therefore, the strength function) and $R^*=R$ in calculations of the partial reaction-amplitudes.
Such a choice of $R^*$ makes the model more consistent in the region of the respective single-particle resonance.
The parameters $\Delta=3$ MeV and $B=7$ MeV in Eq.~(\ref{2}) are taken
to be the same as in Refs.~\cite{Moukh99,Gor00,Rod00},
while parameters $\alpha=0.06$ MeV and $\kappa'$ from Eq.~(2),
are chosen to describe the experimental total width and the energy of the GDR, respectively, in the nuclei
under consideration. The $\kappa'$ values are listed in Table~1:
as can be seen from the latter table, the model parameters used in our
calculations are rather stable.

The quality of description of the experimental photoabsorption cross sections
can be seen in Fig.~1, where the calculated cross sections $\bar \sigma_a(\omega)$
are shown for $^{89}$Y, $^{140}$Ce, $^{208}$Pb target nuclei.
It is worth mentioning that the use of two adjustable parameters $\kappa'$ and $\alpha$
allows one to satisfactorily describe three GDR parameters, namely
the energy, the total width, and the value of $\sigma^{int}_{GDR}$.

The energy-averaged differential partial cross sections for neutron radiative capture,
$d\bar \sigma^{inv}_{\mu}/d \Omega$, are calculated without the use
of any new adjustable parameters. Each calculated cross section is multiplied
by the experimental spectroscopic factor, $S_\mu$, of the final product-nucleus
single-particle  state populated after the capture.
The factors $S_\mu$ are listed in Table~2. The calculated cross sections at $90^0$
and the anysotropy parameters $\bar a_\mu$ are shown in
Figs.~2-4 in comparison with the respective experimental data for the nuclei in question.
We also calculate some partial $^{208}$Pb$(\gamma,n)$-reaction cross sections,
$d \bar \sigma_\mu/d \Omega$, at $90^0$ with the population of single-hole states in $^{207}$Pb
(Fig.~5). The direct-nucleon-decay branching ratios, $b_\mu$, for the GDR in $^{208}$Pb are
calculated according to Eq.~(14) using a spectroscopic factor of unity
for the final single-hole states in $^{207}$Pb. The $b_\mu$ values are listed in Table~3, and the total
branching ratio $b=\sum_\mu b_\mu$ is estimated as 14\%.

To a certain degree, the branching ratios of Eq.~(\ref{eq1}) can be considered to be independent of the GDR
excitation-process. The semimicroscopical approach is also applied to evaluate
the total direct-neutron-decay branching ratio, $b$, for the GDR in $^{48}$Ca:
this value has been deduced from the $^{48}$Ca(e,e$'$n)-reaction cross section~\cite{Str00}.
In the analysis, the E1 strength function, deduced from the (e,e$'$)-reaction in Ref.~\cite{Str00},
is used to determine the parameter $\kappa'$ (Table 1). Unit spectroscopic
factors are also used for the final single-hole states $7/2^-, 1/2^+, 3/2^+$ in $^{47}$Ca.
The calculated value of $b=27\%$ is comparable with the respective experimental
value $b^{exp}=39\pm 5\%$~\cite{Str00}.

\section{Concluding remarks}

As compared to previous attempt of Ref.~\cite{Rod00},
our results allow us to make several comments on capability of the semimicroscopical approach
in describing the simplest photonuclear reactions accompanied by the GDR excitation:
\begin{itemize}
\item[(1)]The use of the "$\omega +\frac i2 I$ method" to take into account phenomenologically
the spreading effect allows us to simplify
the calculations of the energy-averaged reaction cross sections
and also to account for the non-resonant background.
This method can be only used to describe highly-excited giant
resonances within our approach.
\item[(2)] The isoscalar part of separable momentum-dependent
forces is also taken into consideration, and results in difference of the nucleon effective
mass from the free value. It is also seen that the
use of the "realistic" effective mass does not improve the description of the data.
\item[(3)] A reasonable description of
two sets of the rather old experimental data on the partial $(n,\gamma)$-reactions
for medium-heavy mass nuclei~\cite{Berg72,Berg78} is obtained.
Possibly the use of a more elaborate version of the approach
(by taking $(N-Z)/A$ corrections into account,
the use of realistic momentum-dependent forces, etc.) will lead to
better description of the photonuclear reaction data.
\item[(4)] Our consideration of predictions for partial
$^{208}{\rm Pb}(\gamma,n)$-reaction cross section has been
partially motivated by the advent of new facilities,
such as SPring-8 (based on the use of backward Compton scattering),
capable of measuring photonuclear reaction cross sections with high accuracy.
Such electromagnetic probes are more efficient for
studying the nuclear structure. From this point of view the recent experimental
data on the $(e,e'n)$-reaction \cite{Str00} represents a good example.
\end{itemize}

The authors are grateful to Dr.~G.C.~Hillhouse for carefull reading the manuscript
and many useful remarks improving the style.
This work is partially supported by the Russian Fund for Basic Research (RFBR)
under Grant No. 02-02-16655. V.A.R. would like to thank
the Graduiertenkolleg "Hadronen im Vakuum, in Kernen und Sternen" (GRK683)
for supporting his stay in T\"ubingen and Prof.~A.~F\"a{\ss}ler for
his hospitality.

\begin{table}[h]
\caption{Model parameters $U_0$ and $\kappa'$
used in calculations.}
\label{tab1}
\begin{center}
\begin{tabular}{|c|c|c|c|}
\hline
Nucleus & $m^*/m$ &$U_0,$ MeV & $\kappa'$
\\
\hline
$^{89}$Y & 1.0 & 53.3 & 0.53
\\
 & 0.9 & 58.0 & 0.38
\\
\hline
$^{140}$Ce & 1.0 & 53.7 & 0.56
\\
 & 0.9 & 57.9 & 0.39
\\
\hline
$^{208}$Pb & 1.0 & 53.9 & 0.56
\\
 & 0.9 & 58.9 & 0.42
\\
\hline
$^{48}$Ca & 1.0 & 54.3 & 0.48\\
\hline
\end{tabular}
\end{center}
\end{table}
\vskip-1cm
\begin{table}[h]
\caption{Spectroscopic factors of the valence-neutron states
in $^{90}$Y,$^{141}$Ce and $^{209}$Pb (taken from the Ref.~\protect\cite{lik98}).}
\begin{center}
\begin{tabular}{|c|c|c|c|c|c|}
\hline
\multicolumn{2}{|c|}{$^{90}$Y}&\multicolumn{2}{c|}{$^{141}$Ce}
&\multicolumn{2}{c|}{$^{209}$Pb}\\
\hline
$\mu$&$S_\mu$&$\mu$&$S_\mu$&$\mu$&$S_\mu$\\
\hline
$1g_{7/2}$& 0.6 &$2f_{5/2}$& 0.8       &$3d_{3/2}$& 0.9 \\
$2d_{3/2}$& 0.7 &$1h_{9/2}$& 1.0       &$2g_{7/2}$& 0.8 \\
$1h_{11/2}$&0.4&$1i_{13/2}$& 0.6      &$4s_{1/2}$& 0.9 \\
$3s_{1/2}$& 1.0 &$3p_{1/2}$& 0.4       &$3d_{5/2}$& 0.9 \\
$3d_{5/2}$& 1.0 &$3p_{3/2}$& 0.4       &$1j_{15/2}$& 0.5\\
          &     &$2f_{5/2}$& 0.8       &$1i_{11/2}$& 1.0\\
          &     &          &           &$2g_{9/2}$& 0.8 \\
\hline
\end{tabular}
\end{center}
\end{table}

\begin{table}[h]
\caption{Calculated branching ratios for direct neutron decay
of the GDR in $^{208}$Pb. }
\label{tab3}
\begin{center}
\begin{tabular}{|c|c|c|c|c|c|c|c|}
\hline
 & $m^*/m$ & \multicolumn{6}{|c|}{Final single-hole states}\\
\cline{3-8}
& &3p$_{{\rm \frac 12}}$ &2f$_{{\rm \frac 52}}$& 3p$_{{\rm \frac 32}}$ &
1i$_{{\rm \frac{13}2}}$ &2f$_{{\rm \frac 72}}$& 1h$_{{\rm \frac 92}}$\\
\hline
$b_\mu$,& 1.0 & 2.0&4.4&3.4&1.3&2.5&0.7\\
 \%     & 0.9 &1.8 &3.6&3.1&1.0&1.8&0.4\\
\hline
\end{tabular}
\end{center}
\end{table}

\newpage
\begin{figure}
\centerline{\epsfig{file=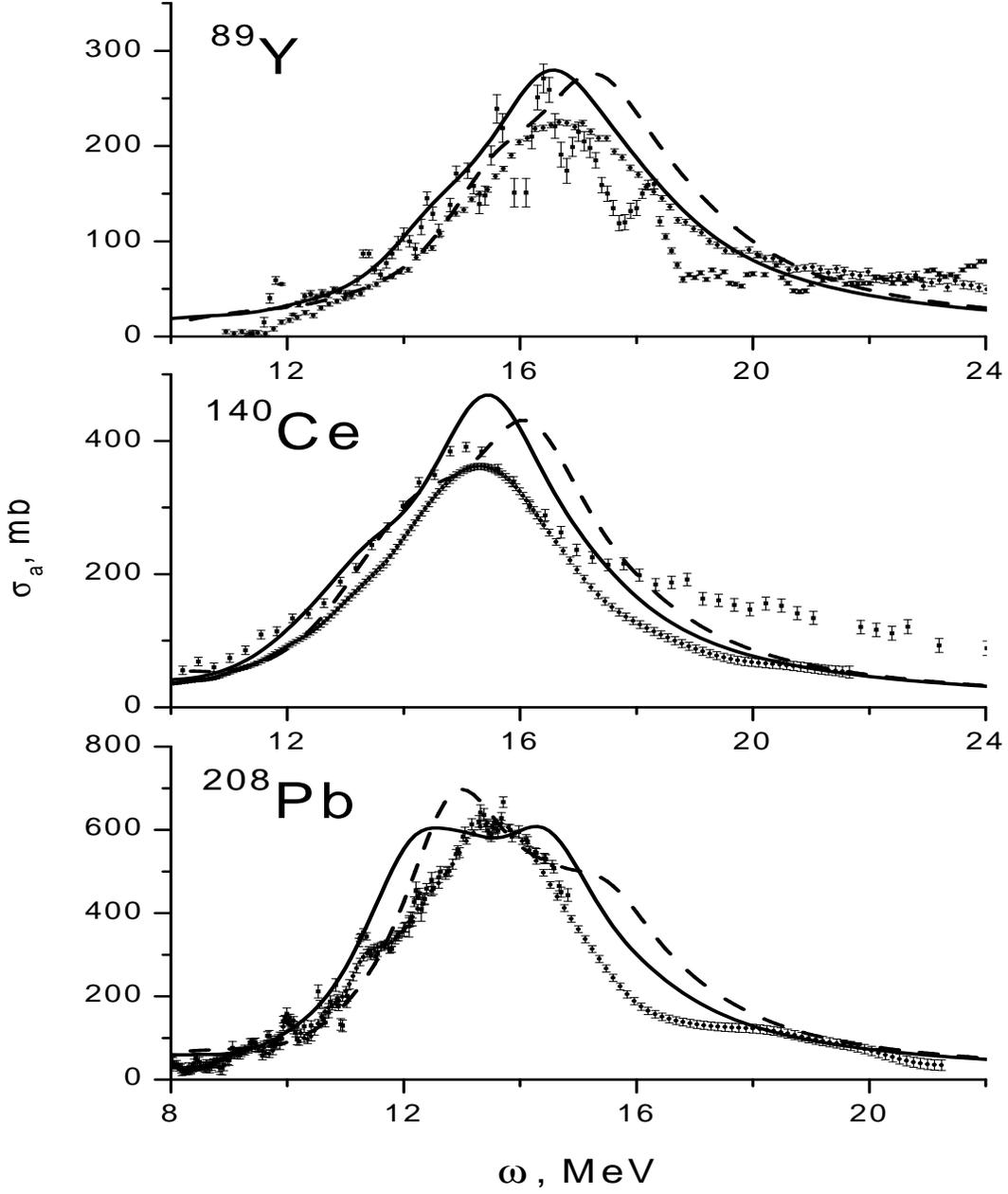,height=7.in,width=6.in}}
\vspace{10pt}
\caption{The calculated photoabsorption cross sections, $\bar\sigma_a$,
for $^{89}$Y, $^{140}$Ce and $^{208}$Pb (hereafter the solid and dashed lines correspond
to calculations with $m^*/m=\ 1$ and 0.9, respectively).
The experimental data (black squares and circles) are taken from
Refs.~\protect\cite{Ber75}.}
\label{fig1}
\end{figure}
\newpage
\begin{figure}
\vspace{30pt}
\centerline{\epsfig{file=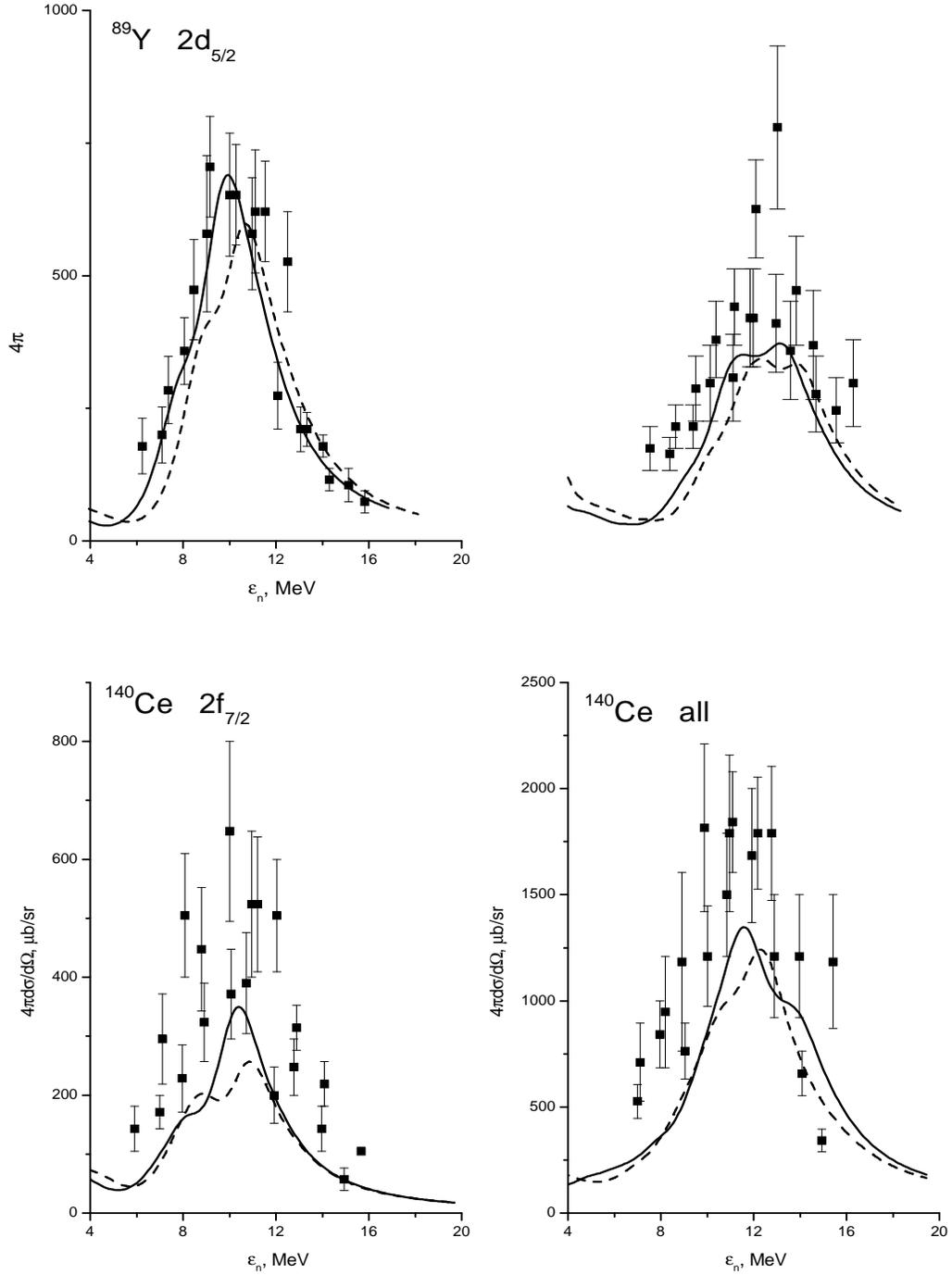,height=8.in,width=6.in}}
\caption{The calculated partial cross sections at $90^0$ multiplied
by $4\pi$ as functions of neutron
energy for neutron radiative capture to the ground state
and to all the
single-particle states in $^{90}{\rm Y}$ and $^{141}{\rm Ce}$.
The experimental data are taken from Ref.~\protect\cite{Berg78}.}
\label{vrod:fig1}
\end{figure}
\newpage
\begin{figure}[htb]
\vspace{30pt}
\centerline{\epsfig{file=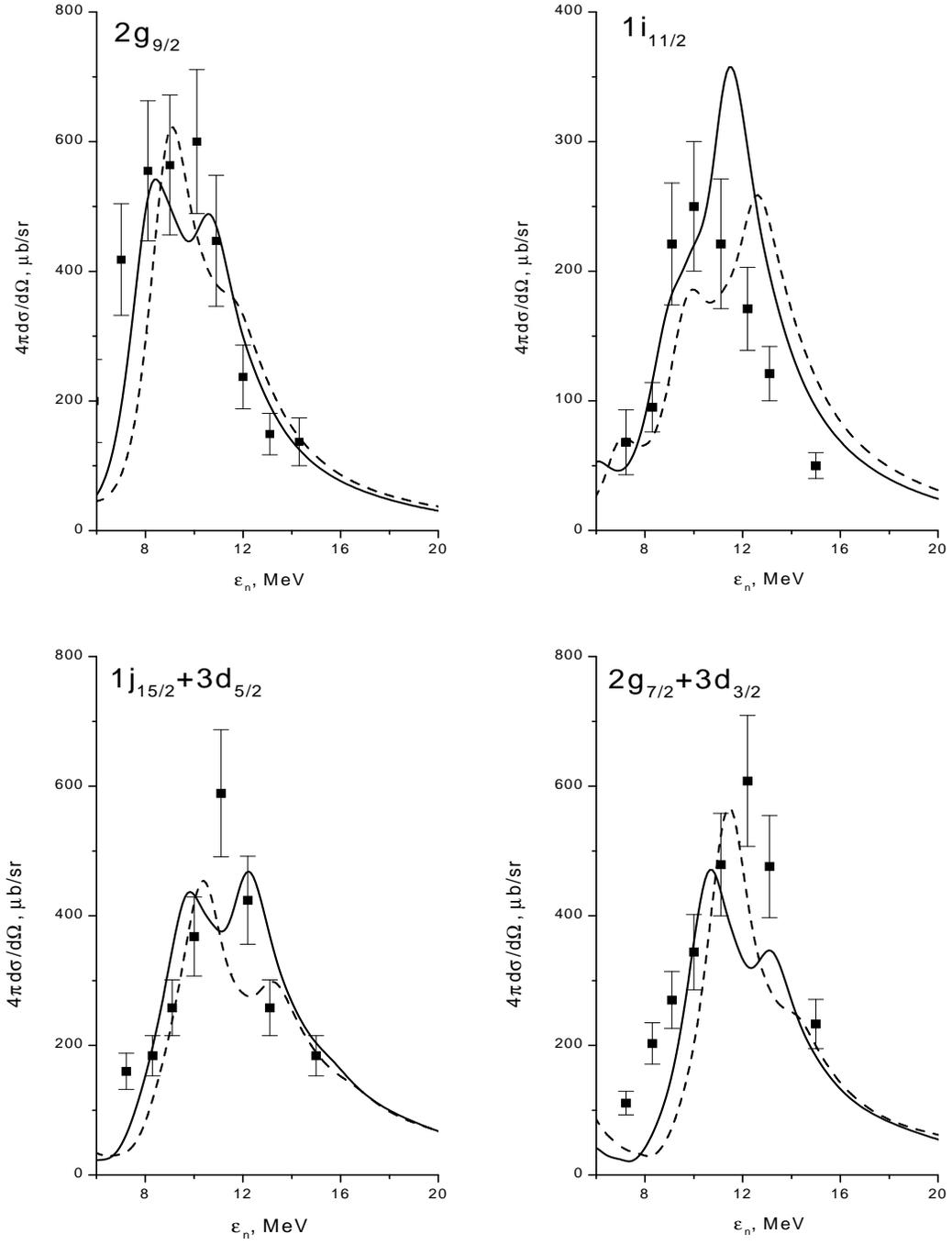,height=8.in,width=6.in}}
\caption{The calculated partial cross sections at $90^0$ multiplied
by $4\pi$ as functions of neutron energy for neutron radiative capture
to some single-particle states in $^{209}{\rm Pb}$.
The experimental data are taken from Ref.~\protect\cite{Berg72}.}
\label{fig3}

\end{figure}
\newpage
\begin{figure}[htb]
\vspace{30pt}
\centerline{\epsfig{file=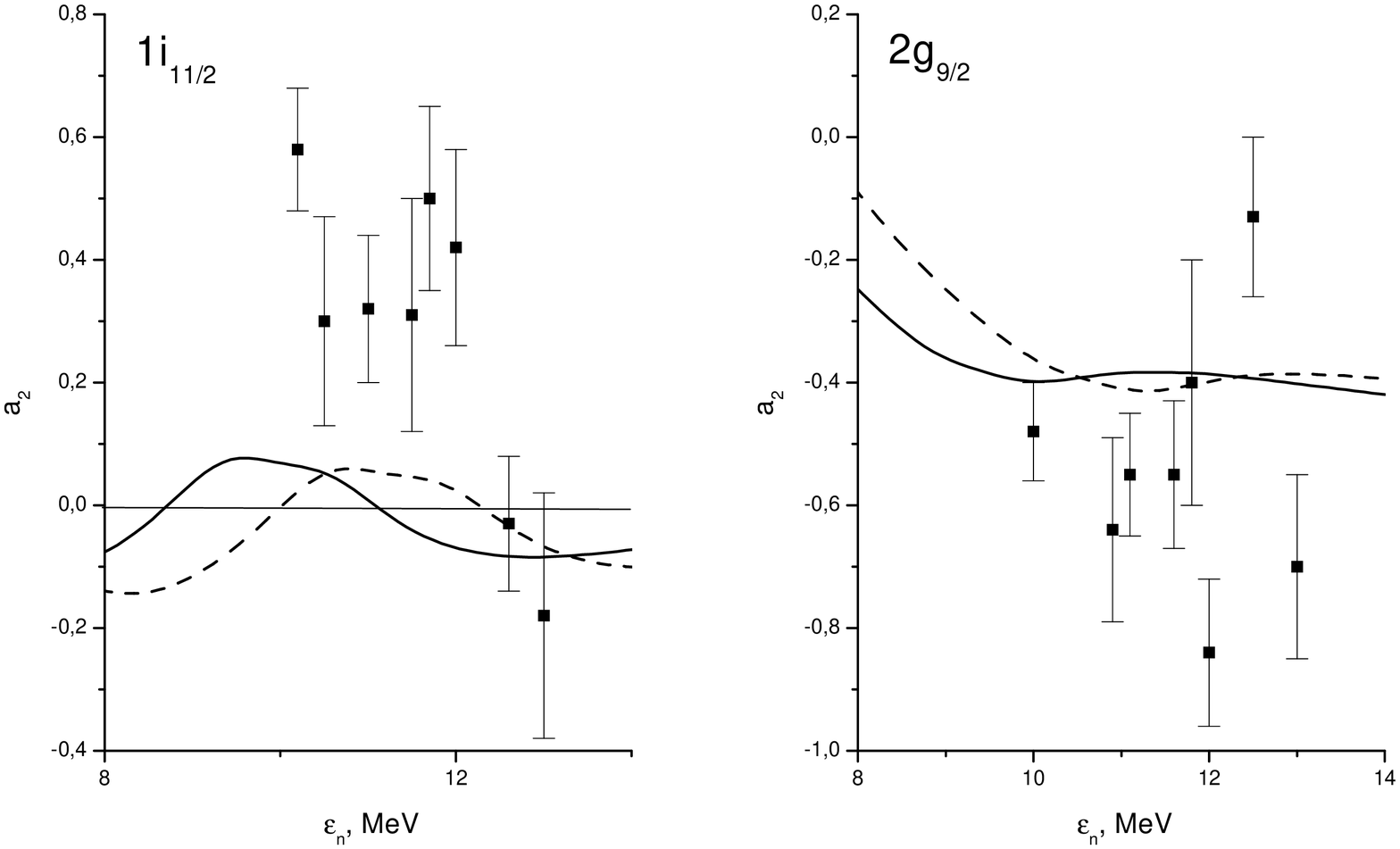,height=4.in,width=6.in}}
\caption{ Calculated anisotropy parameter $a_2$ for some partial
$^{208}{\rm Pb}(n,\gamma)$-reactions. The experimental data are taken
from Ref.~\protect\cite{Berg72}.}
\label{fig4}
\end{figure}
\newpage
\begin{figure}[htb]
\vspace{30pt}
\centerline{\epsfig{file=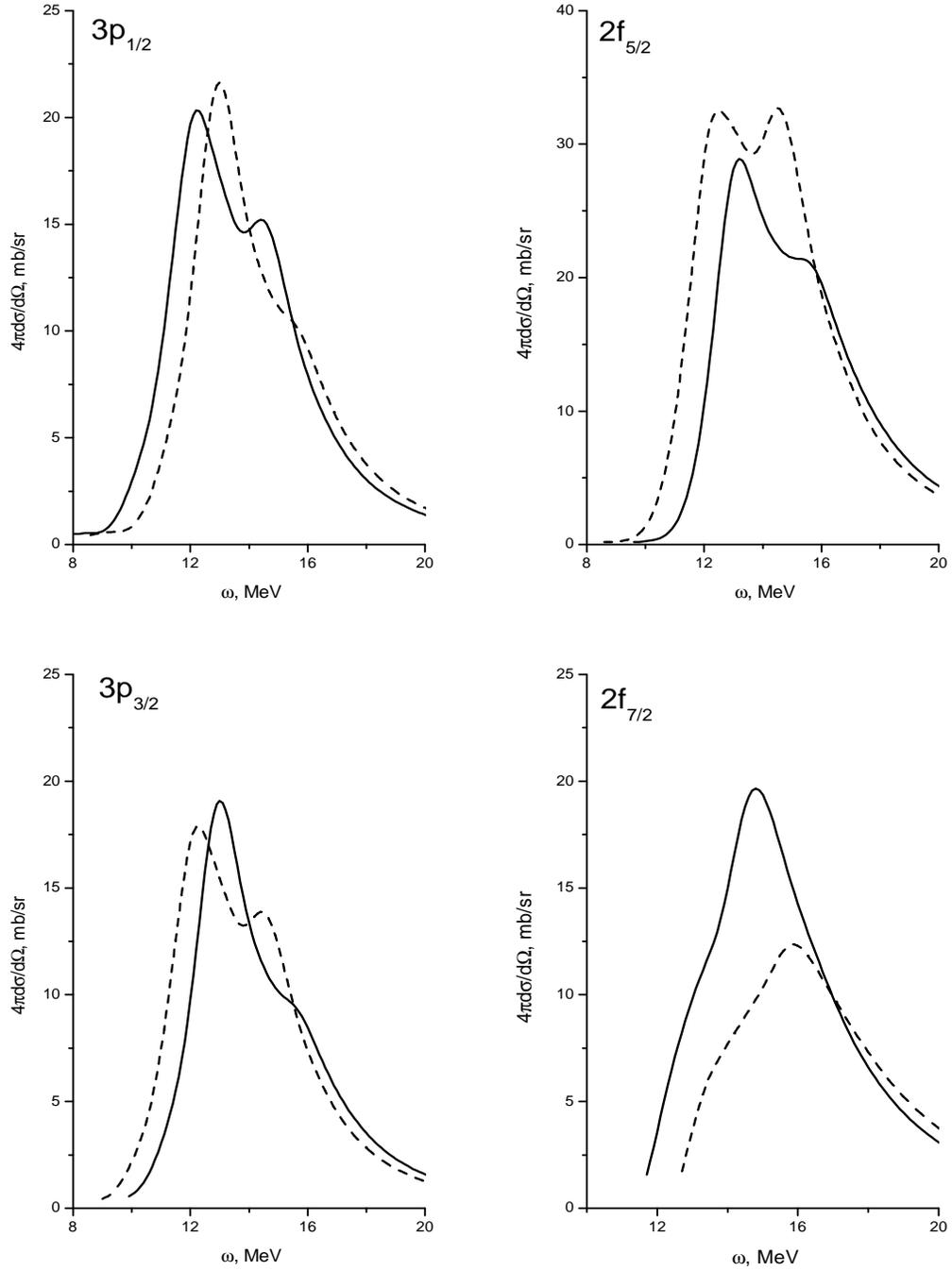,height=8.in,width=6.in}}
\caption{The calculated partial cross sections at $90^0$ multiplied by $4\pi$
of the $(\gamma,n)$-reaction
with population of some single-hole states in $^{207}{\rm Pb}$
as functions of photon energy.}
\label{fig5}
\end{figure}
\end{document}